\documentclass[a4paper]{JACoW-GSI}

\usepackage[tbtags]{amsmath}
\usepackage{amssymb}
\usepackage{amsfonts}
\usepackage{euscript}
\usepackage{longtable}
\usepackage[dvips]{graphicx}
\usepackage{multicol}
\usepackage{color}
\usepackage{rotating}
\usepackage{hhline}
\usepackage[matrix,arrow,curve]{xy}

\begin{document}

\title{A strong laser impact on spin precession of a charged particle  in
the semi-relativistic interaction regime}

\author{A.~Khvedelidze\,${}^{a,b,c}$\, and  D. Mladenov \,${}^{d}$}

\affil{${}^a$  {\it Laboratory of Information Technologies,
Joint Institute for Nuclear Research, Dubna,
Russia} }

\affil{ ${}^b$
{\it A. Razmadze Mathematical Institute,
Tbilisi State University,
Tbilisi, Georgia}}

\affil{${}^c$ \it  School of Natural Sciences,  University of Georgia, Tbilisi, Georgia}

\affil{${}^d$ \it Faculty of Physics, Sofia State University, Sofia, Bulgaria}

\date{\empty}

\maketitle

{$\bullet$ \bf Motivation and main results $\bullet$}
With this note  we aim to report on studies of classical and  quantum nonlinear effects
in dynamics of a charged particle's spin  caused  by the interaction with a high intensity laser  modeled by the monochromatic plane wave.

The standard dipole approximation \cite{Jackson} in laser-matter interaction is well justifiable
and consistent with the non-relativistic treatment of a charged particle dynamics if the laser intensity is low.
With the growing intensity the retardation effects as well as the influence of the magnetic part of the
Heaviside-Lorentz force become important and modification of the dipole approximation schemes becomes unavoidable.

Quantity characterizing the intensity of a coherent radiation
interactions with a massive $(m)$, charged particle $(e)$
is given by the so-called laser field strength parameter:
\[
\eta^2: =-2\,\frac{e^2}{m^2c^4}\,\langle \langle\, {A_\mu
A^\mu}\,\rangle\rangle\,,
\]
where $\langle \langle \cdots \rangle\rangle$ denotes time average
and  $ A_\mu \,$ is a four vector describing the  laser field~\cite{Sengupta,SarachikSchappert}.
It is well established that for a vanishing  field strength $\eta^2 \ll 1,$
a charged particle's spin behaves adiabatically, linearly
responding to the  magnetic component of the electromagnetic field.
When $\eta^2 $ is increasing the dipole approximation breaks down, non-linear effects come in force and
the adiabatic picture of the spin precession turns to be inadequate.
Below, having in mind the necessity of the dipole approximation scheme modification,
we briefly state a method  allowing to describe the  evolution of a spin-1/2 particle
in  a strong monochromatic plane  wave for  the semi-relativistic interaction regime.
Particularly,  the question what is the effect of
complete treatment of the Heaviside-Lorentz force (non ignoring
its magnetic part) on the solution to the Larmor precession is addressed.
Our consideration includes both  classical as well as the
non-relativistic quantum mechanical analysis.
The results stated below developed  out of our studies of a charged particle's classical dynamics
beyond the dipole approximation \cite{JamesonKhvedelidze}
and  semiclassical analysis of the  spin-1/2 evolution in a strong laser field
\cite{EliashviliGerdtKhvedelidze,GerdtGogilidzeKhvedelidzeMladenovSanadze}.
Our description of the spin dynamics  is applicable  in the semi-relativistic  regime of light-matter  interactions,
when  $\eta^2$ is not negligible any more,  but still $\eta^2< 1$  and  expansions over $\eta^2$  are correct.

Employing the above methods, the following effects of a strong coherent radiation have been found:
\begin{enumerate}
  \item A nonlinear dependence of the spin precession
  frequency on a laser intensity and polarization;
  \item  Appearance of the quantum phase of spin's  wave function due to the nonvanishing laser intensity.
\end{enumerate}
{$\bullet$ \bf Background and calculations $\bullet$}
We follow the semiclassical attitude towards the spin dynamics;
spin degrees of a charged particle are described in a standard quantum mechanical way using
the Pauli equation with the leading relativistic corrections.
The effective spin-radiation interaction has a conventional form with magnetic field evaluating
along charged particle's classical trajectory found from Newton's
equations beyond the dipole approximation \cite{JamesonKhvedelidze}.

The basic assumption on a charged particle state vector
$|\Psi\rangle $
consists in charge $\&$ spin decomposition:
\footnote{Here to simplify the expressions, we keep only one nonzero coefficient, $c_{+,0}\,.$
The unit normalization condition on the WKB wave function fixes this
coefficient, $\pi c^2_{+,0} = 2m\,\omega_{P}\,,$  where $\omega_{P}$
is fundamental frequency of a particle driven by a laser.}
\begin{equation}\label{eq:spin-charge.decomp}
|\Psi\rangle = \sum_{i= 0,1}\,\sum_{\alpha=\pm}\ c_{\alpha,
i}|\psi_\alpha\rangle\otimes|\chi_i\rangle\,.
\end{equation}
Two states $|\psi_\pm\rangle\,,$ are linearly independent WKB
solutions to the Schr\"{o}dinger equation for a charge
interacting with a laser radiation in the radiation gauge,
$A_0 =0, \ \boldsymbol{\nabla}\cdot\mathbf{A}=0$.
The vector $|\chi_i\rangle\,$  describes the states of a non-relativistic spin-1/2
subject to the Pauli equation with an effective Hamiltonian derived in the semiclassical approximation from the
Schr\"{o}dinger equation.
Our calculations lead to the following Hamiltonian operator
governing the evolution of a two component spin-1/2 wave function
$|\chi(t)\rangle$:
\begin{equation*}\label{eq:Hanmiltoer}
    H(t):= -\frac{\kappa\hbar }{2}\,\boldsymbol{\Omega}(t)
    \cdot\boldsymbol{\sigma}\,.
\end{equation*}
Here $\kappa=e/mc $ and  $\boldsymbol{\sigma}$ are the Pauli matrices.
The Larmor vector $\boldsymbol\Omega(t)=(\Omega_1, \Omega_2,  \Omega_3)$
depends on time by means of the Jacobian elliptic  functions,
whose argument is $u=\omega^\prime t $ and the modulus reads
$\mu^2=(1-2\varepsilon^2)\eta^2\gamma_z$:
\begin{eqnarray}
&&
{\Omega}_1=\eta\omega^\prime \varepsilon^\prime\mathrm{sn}(u,\mu)\mathrm{dn}(u, \mu)\,,\\
&&
{\Omega}_2=\eta\omega^\prime \varepsilon\,\mathrm{cn}(u, \mu)\mathrm{dn}(u,\mu)\,,\\
&&
{\Omega}_3=- \eta^2\varepsilon\varepsilon^\prime{\omega}\,,
\end{eqnarray}
where $\varepsilon $  denotes a laser polarization,  $\varepsilon^\prime=\sqrt{1-\varepsilon^2}$  and
the $\omega^\prime\gamma_z=\omega$  stands for the non-relativistically Doppler shifted
laser frequency $\omega$ with a contraction factor fixed by particle's $ z$-component velocity, $\gamma_z^{-1}=1-v_z(0)/c$.
Our analysis of the spin evolution equation for two extreme polarizations,
 \textit{linear} and  \textit{circular},  reveals  the following pattern.

\bigskip

\centerline{\bf Linear polarization}
\bigskip

In this case the spin-1/2 wave function
exposes nontrivial  periodicity:
\begin{equation}\label{eq:linpol}
|\chi(t+4\mathrm{K}(\mu )/\omega^\prime )\rangle=|\chi(t)\rangle\,,
\end{equation}
where
$\mathrm{K}(\mu )$ is the  quarter period  of the  Jacobian function.
Apart from the pure kinematical
non-relativistic Doppler shift  the particle's  spin
precession frequency depends on the laser intensity
through the period $\mathbb{K}$.
Furthermore, a spin precesses at frequency that depends nonlinearly on the laser intensity.
For the semi-relativistic intensities $\eta \ll 1\,,$ the period of the particle oscillation can be
represented in the form of the expansion
\begin{equation*}\label{eq:smallfreq}
T_P=\frac{2\pi}{\omega^\prime}\left[
1+\left(\frac{1}{2}\right)^2{1-2\,\varepsilon^2}{\gamma_z^2}\,\eta^2
+
\dots
\right].
\end{equation*}
The presence of $\mathbb{K}$ in (\ref{eq:linpol}) exposes  a fundamental
peculiarity of the particle dynamics which is beyond the dipole approximation.

\bigskip

\centerline{\bf  Circular polarization}
\bigskip

For a particle interaction with a circularly polarised laser there
is no modification of the precision period caused by a laser intensity,
but another, pure quantum mechanical effect occurs.
The integration of the spin precession equation gives
\begin{equation}\label{eq:solunit}
    |\chi(t)\rangle= U(0,t)\,|\chi(0)\rangle\,, \qquad U(0,0)= I\,,
\end{equation}
where the evolution operator  $U(0,t)$ in the Eulerian form reads
\begin{eqnarray*}
 U(0,t)&=&e^{\displaystyle -i\omega^\prime t\,\frac{\sigma_3}{2}}
\cdot\\
&& e^{\displaystyle -i\beta\,\frac{\sigma_1}{2}}\,
    e^{\displaystyle-i\eta\kappa\omega^\prime\sqrt{1+\Delta^2} t \frac{\sigma_3}{2}}
  e^{\displaystyle i\beta\,\frac{\sigma_1}{2}}\,,
\end{eqnarray*}
where \( \tan\beta:= \Delta\,,\  \Delta:=
 - 1/{\kappa\eta}+ \eta^2\gamma_z\,.
 \)
From this solution we conclude that the precession period is not modified but the state changes under the cyclic evolution highly  nontrivially.
The spin evolution operator $U(0, t)$ undergoes the following change under the cyclic evolution
\[
U\left(t+{2\pi}/{\omega^\prime}\right)=e^{i\pi} U\left(t\right)
M\,,
\]
with the so-called \textit{monodromy} matrix
\begin{equation}\label{eq:monodromymatrix}
    M(\omega^\prime )=
  e^{\displaystyle -i\beta\,\frac{\sigma_1}{2}}\,
   e^{\displaystyle i\pi\eta\kappa\sqrt{1+\Delta^2}\sigma_3}
   \,e^{\displaystyle i\beta\,\frac{\sigma_1}{2}}
    \,.
\end{equation}
The monodromy matrix $M$ depends explicitly on the  laser intensity.
If the special initial conditions on the state vector are chosen the monodromy matrix (\ref{eq:monodromymatrix}) diagonalises
 \[
 M_D= e^{\displaystyle i\pi\eta\kappa\sqrt{1+\Delta^2}\,\sigma_3}.
 \]
Note, that  for the vanishing laser field strength, the induced quantum phase reduces to its non-relativistic value.

{$\bullet$ \bf Conclusion $\bullet$}
Our studies  point on new effects associated with a
spin-1/2 charged particle motion in laser fields with intensities
corresponding to the semi-relativistic interaction regime.
These  effects are identified beyond the dipole approximation
and  could be attributed to the fact that the magnetic forces associated with the laser field have been retained and  alter the
spin evolution appreciably.

{$\bullet$ \bf Acknowledgments $\bullet$ \ }
Work is supported in part by the JINR-Bulgaria collaborative grant,
the University of Georgia,
the ICTP-SEENET-MTP Network project PRJ-09 ``Cosmology and Strings``
and  the Ministry of Education and
Science of the Russian Federation (grant No. 3802.2012.2).


\end{document}